\documentclass[a4paper]{article}
\usepackage[utf8]{inputenc}
\usepackage{geometry}
\usepackage{hyperref}
\usepackage{url}
\usepackage{amsmath}
\usepackage{amsthm}
\usepackage{amssymb}
\usepackage{mathtools}
\usepackage{graphicx}
\usepackage{textcomp}
\usepackage{microtype}
\usepackage{authblk}
\usepackage{multicol}
\usepackage[table,xcdraw]{xcolor}
\usepackage{float}
\providecommand{\keywords}[1]
{
  \small	
  \textbf{\textit{Keywords---}} #1
}

\title{An anatomy-based V1 model: Extraction of Low-level Features,  Reduction of distortion and a V1-inspired SOM}
\author[1,*]{Suvam Roy}
\author[2,*]{Nikhil Ranjan Pal}
\affil[1]{Department of Physical Sciences, Indian Institute of Science Education and Research Kolkata, Mohanpur-741246, India}
\affil[2]{Electronics and Communication Sciences Unit, Indian Statistical Institute, Baranagar-700108, India}
\affil[*]{Address for correspondence: \href{mailto:rsuvam1996@gmail.com}{rsuvam1996@gmail.com}, \href{mailto:nikhil@isical.ac.in}{nikhil@isical.ac.in}}

\date{\today}

\begin{document}

\maketitle


\begin{abstract}
    We present a model of the primary visual cortex V1, guided by anatomical experiments. Unlike most machine learning systems our goal is not to maximize accuracy but to realize a system more aligned to biological systems.  Our model consists of the V1 layers 4, 2/3, and 5, with inter-layer connections between them in accordance with the anatomy. We further include the orientation selectivity of the V1 neurons and lateral influences in each layer. Our V1 model, when applied to the BSDS500 ground truth images (indicating LGN contour detection before V1), can extract low-level features from the images and perform a significant amount of distortion reduction. As a follow-up to our V1 model, we propose a V1-inspired self-organizing map algorithm (V1-SOM), where the weight update of each neuron gets influenced by its neighbors. V1-SOM can tolerate noisy inputs as well as noise in the weight updates better than SOM and shows a similar level of performance when trained with high dimensional data such as the MNIST dataset. Finally, when we  applied V1 processing to the MNIST dataset to extract low-level features and trained V1-SOM with the modified MNIST dataset, the quantization error was significantly reduced. Our results support the hypothesis that the ventral stream performs gradual untangling of input spaces.
\end{abstract}

\vspace{0.3cm}

\vspace{0.4cm}

\begin{multicols}{2}


\section{Introduction}
\paragraph{}
Deep learning has witnessed tremendous progress in the last decade. In the case of visual object recognition, Convolutional Neural Networks (CNNs) have exhibited great success  \cite{Alexnet,VGGnet,Googlenet,Resnet,Imagenet}. However, along with its success, deep learning models have deviated significantly from the way the brain performs object recognition. Earliest deep learning models, such as the Neocognitron, \cite{Neocognitron} were inspired by the anatomical information available at that time about the brain, like the presence of simple cells and complex cells in the primary visual cortex \cite{Hubel_1962,Hubel_1965}. Since then experiments have revealed the presence of more complex structures in the brain. Deep learning however did not follow these newer biological developments.  Another major problem of deep learning is its lack of transparency and explainability. Moreover, the solutions obtained after training on a dataset are often sensitive to the initial conditions, implying that no causal relationship has been identified by the net to arrive at the  set of solutions. We believe that there is a need to develop visual recognition systems that share more attributes of biological visual systems. One might ask what is the need for developing bio-realistic learning models when the alternative route is performing quite well! In our view, such systems may help to achieve better transparency. In the recent past, some researchers have strongly emphasized the need of exploiting the knowledge of the pre-frontal cortex, in designing deep learning systems \cite{bengio_2020,bengio_2022}. In fact, there have been few attempts at developing biology-inspired learning models. One example is a reward-based learning model BrainProp \cite{Brainprop}. However, this model still uses a CNN architecture and category labels that are not consistent with knowledge of biological systems. Another bio-inspired model is the CORnet-S which uses a complete ventral stream architecture (V1, V2, V4, ITC) for image recognition \cite{Cornet_S}. The internal structure of each of the 4 regions actually contains a CNN architecture as well. Therefore, these attempts though remarkable, none of them makes proper use of the recent findings of neuroscience experiments. 

\begin{figure*}[!ht]
    \centering
    \includegraphics[width=\textwidth]{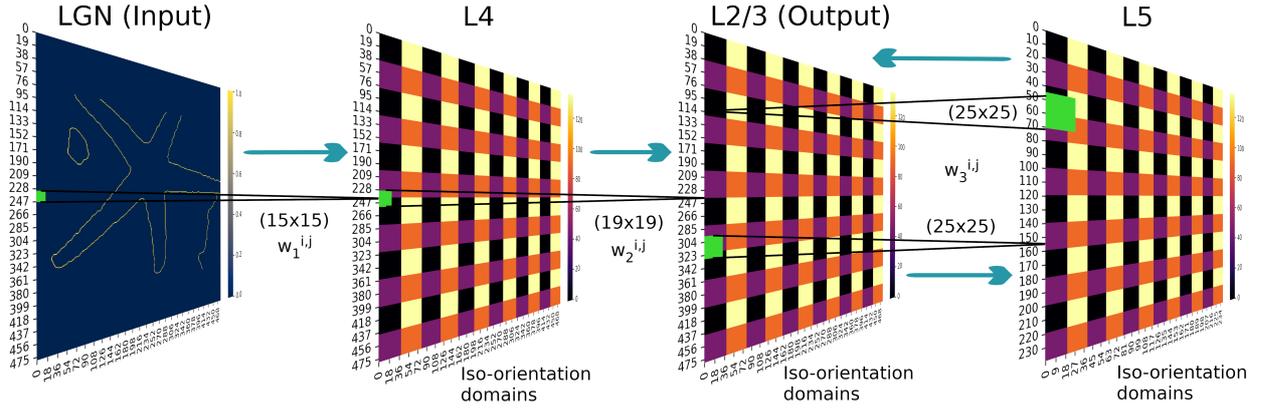}
    \caption{V1 model architecture}
    \label{fig_1}
\end{figure*}

\begin{figure*}[!ht]
    \centering
    \includegraphics[width=0.8\textwidth]{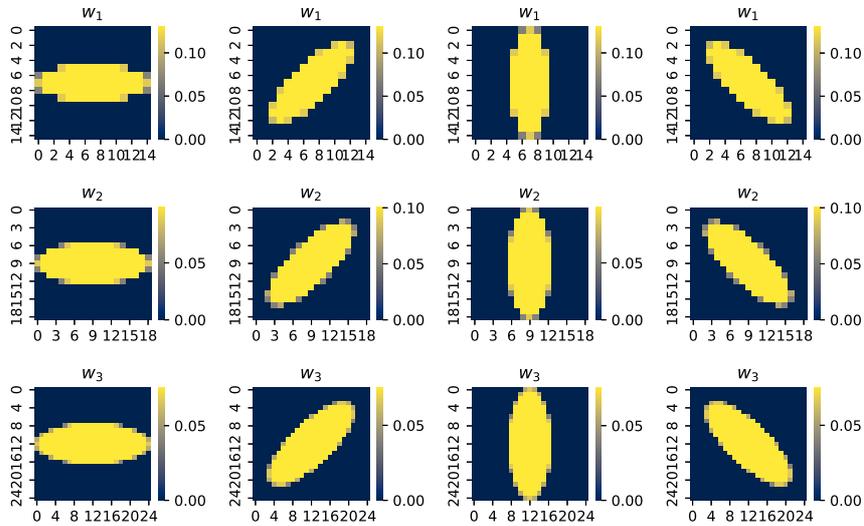}
    \caption{Afferent weights between different V1 layers at different iso-orientation columns. 1st row: between LGN and L4; 2nd row: between L4 and L2/3; 3rd row: between L2/3 - L5 and L5 - L2/3.}
    \label{fig_2}
\end{figure*}

\paragraph{}
In this work, we try to model the first stage of the ventral stream, i.e., the primary visual cortex V1, by taking inspiration from recent anatomical experiments. In our model, we have taken into consideration the neuron and synapse densities in the V1 layers, the orientation selectivity of the V1 neurons, as well as the lateral influences resulting from the interplay of pyramidal and interneurons. All of these biological details have been incorporated into our model. However, to keep our model simple, we do not go to the level of details considered in Spiking Neural Network models \cite{spiking_neuron}.  We consider 4 transitions between LGN (Lateral Geniculate Nucleus) and V1 layers in our model: LGN to L4, L4 to L2/3, L2/3 to L5, and L5 to L2/3. When this V1 model acts on LGN contour-detected inputs, it can extract low-level features from those inputs. Our V1 model can also perform a significant amount of distortion reduction when distorted versions of the inputs are presented. 

\paragraph{}
Inspired by our V1 model we also propose a modified Self-organizing map (V1-SOM) algorithm. The  V1-inspired model tries to computationally exploit the role of V1 in object recognition. In the original self-organizing map algorithm \cite{Kohonen_som} the weights of different neurons actually get updated independently  depending on the spatial distance from the winner neuron. In our V1-SOM the weight update of every neuron gets influenced by the neighboring neurons. V1-SOM shows better tolerance to noisy inputs. It also exhibits robustness to perturbation of weights during learning of 2D distributions and shows similar performance on the MNIST data \cite{mnist}. Interestingly when we apply the V1 model to the MNIST dataset and use the V1 processed MNIST data as inputs, the quantization error gets reduced significantly for both SOM and V1-SOM.

\paragraph{}
We also show that the performance of this more biologically plausible self-organizing map algorithm is equally good as the traditional SOM algorithm. Our  results obtained from V1 and V1-SOM are supportive of the \textit{Manifold unfolding hypothesis} of the ventral stream \cite{Manifold_hypothesis}, according to which each region of the ventral stream causes gradual untangling of the input manifolds, to separate out objects belonging to different classes, while bringing those belonging to the same class closer to each other.

\section{Methods}
\subsection{V1 model}

\paragraph{}
LGN cells in the thalamus along with early primary visual cortex (V1) cells have been shown to perform contour detection from retinal input images \cite{Corf_2012, Corf_2014,vanrullen}. In our model, we use such post-LGN contour-only images as inputs to the V1. LGN transmits most of its outputs to layer 4 of V1, while V2 receives most of its input from layer 2/3 of V1 \cite{Monkey_layers}. Therefore, layer 4 and layer 2/3 are the input and output layers of the V1. Within the different layers of V1 there are strong excitatory connections from L4 to L2/3, from L2/3 to L5 and back from L5 to L2/3 \cite{Monkey_layers}. Based on this we design a 4-layer V1 model consisting of the following transitions: LGN to L4, L4 to L2/3, L2/3 to L5, and L5 to L2/3 (Figure-\ref{fig_1}). We take the dimensions of L4 and L2/3 layers equal ($m\times n$), while for the L5 layer, we take the dimension to be half of the dimensions of L4 and L2/3 layers, i.e., ($m/2\times n/2$), following experimental neuronal densities \cite{LGN_to_L4}. Now according to the neuroscience experiment \cite{LGN_to_L4}, each L4 neuron gets input from $\sim 200$ LGN neurons. Hence we connect each neuron of the L4 layer with ($15\times 15$) LGN neurons around its direct perpendicular location. In other words, the $(i, j)^{th}$ neuron in L4 is connected to LGN neurons with indices varying from (i-7) to (i+7) and (j-7) to (j+7).  Each L2/3 neuron gets excitatory input from $(\sim 300 - 400)$ L4 neurons \cite{L4_to_L23}. We, therefore, connect each L2/3 neuron to ($19\times 19$) L4 neurons perpendicular to it. Finally, comparing the projection fields of L4 neurons in L2/3 layer and L2/3 neurons in L5 layer \cite{L4_to_L23} and based on strong bi-directional connections between L2/3 and L5 layers \cite{Monkey_layers}, we estimated that each L5 neuron gets input from $\sim 600$ L2/3 neurons and vice versa. We, therefore, connect each L5 node with ($25\times 25$) L2/3 nodes and each L2/3 node with ($25\times 25$) L5 nodes around the direct perpendicular locations. Next, we include the existence of iso-orientation columns across all V1 layers, which are found in cats and all other higher cognitive animals \cite{Orientation_map_1,Orientation_map_2,Orientation_map_3,Orientation_map_4}. We assume that each neuron in each layer is selective to 1 out of 4 angles of orientation/directions ($0^\circ, 45^\circ, 90^\circ, 135^\circ$), depending on the orientation domain it belongs to. We take the size of iso-orientation domains ($40\times 40$) for layers 4 and 2/3, and ($20\times 20$) for layer 5. Figure-\ref{fig_2} shows all of the possible inter-layer weight matrices according to the orientation selectivity maps. Finally, we also consider the effect of lateral connections in each layer. All of the V1 layers considered in our model contain $\sim 80\%$ excitatory pyramidal neurons and $\sim 20\%$ inhibitory interneurons \cite{Monkey_layers}. While inter-layer connections are mostly excitatory connections, intra-layer connections on the other hand are mostly inhibitory type \cite{Monkey_layers,Layer_4,Layers,Layer_23}, because the range of lateral arborization in the case of pyramidal neurons is much smaller compared to interneurons. As a result, the lateral influence of a pyramidal neuron that receives excitatory input from the previous layer has a small range. However long-range lateral influence by pyramidal neurons is still possible by an indirect 2-step process, where a pyramidal neuron first excites an interneuron, which in turn inhibits another neuron further away. Such disynaptic inhibition can give rise to Mexican hat-shaped lateral influences \cite{Mexican_hat_from_anatomy}. Based on these above-mentioned findings we consider an approximate scenario in our model. We assume all nodes are excitatory type, as most neurons are pyramidal neurons in the L4, L2/3, and L5 layers. However, we also consider the effect of interneurons by considering Mexican hat-like lateral influences by each pyramidal neuron in all of the layers.   

\paragraph{}
At each layer, we consider a 2 step process for activity generation. In the 1st step, the neurons receive inputs from the previous layer and in the 2nd step, the neurons influence each other through lateral connections. The activity at layer 4 in the first step can therefore be expressed as,

\begin{equation}
    L4_{i,j}=\sum_{k=i-7}^{i+7} \sum_{l=j-7}^{j+7} w_1^{i,j,k,l} LGN_{k,l}
    \label{equ_1}
\end{equation}

\paragraph{}
where $L4_{i,j}$ and $LGN_{k,l}$ denote the activity of the neuron at location ($i,j$)  on layer 4 and of the neuron at location ($k,l$)  on the LGN layer, with $w_1^{i,j,k,l}$ denoting the connection weight between those two nodes. The limits of the summation ensure the use of a $15 \times 15$ neighborhood.

\paragraph{}
In the lateral influence step, each neuron spreads its activity to its neighbors. We assume that the amount of activity received by a neuron at location (i,j) from a neuron at location (k,l) is given by the Marr wavelet (which creates 2D Mexican hat distribution that mimics the effect of disynaptic inhibition as mentioned in \cite{Mexican_hat_from_anatomy}),

\begin{equation}
    M_{\sigma}(i,j;k,l)=\left [ 1-\frac{(i-k)^2+(j-l)^2}{2\sigma^2} \right ] e^{-\frac{(i-k)^2+(j-l)^2}{2\sigma^2}}
    \label{equ_2}
\end{equation}

\paragraph{}
We normalize the Marr wavelet influence originating from each neuron (i,j) as $M_{\sigma}(k,l;i,j)=\frac{M_{\sigma}(k,l;i,j)}{\sum_{k,l}M_{\sigma}(k,l;i,j)}$. After  step 2, the activity of each L4 neuron becomes,

\begin{multline}
    L4_{i,j}^{\prime}=L4_{i,j}\left [ 1-\sum_{k,l}M_{\sigma}(k,l;i,j) \right ] \\ + \sum_{k,l} L4_{k,l} M_{\sigma}(i,j;k,l)
    \label{equ_3}
\end{multline}

\paragraph{}
Similar to Equations \ref{equ_1} and \ref{equ_3}, for L4 to L2/3 and L2/3 to L5 transitions the activities can be expressed as in Equations \ref{equ_4} and \ref{equ_5}, respectively.

\begin{equation}
    L2/3_{i,j}=\sum_{k=i-9}^{i+9} \sum_{l=j-9}^{j+9} w_2^{i,j,k,l} L4_{k,l}^{\prime}
    \label{equ_4}
\end{equation}

\begin{multline}
    L2/3_{i,j}^{\prime}=L2/3_{i,j}\left [ 1-\sum_{k,l}M_{\sigma}(k,l;i,j) \right ] \\ + \sum_{k,l} L2/3_{k,l} M_{\sigma}(i,j;k,l)
    \label{equ_5}
\end{multline}

\begin{equation}
    L5_{i,j}=\sum_{k=i-12}^{i+12} \sum_{l=j-12}^{j+12} w_3^{i,j,k,l} L2/3_{k,l}^{\prime}
    \label{equ_6}
\end{equation}

\paragraph{}
In the lateral influence step of layer 5, we take the $\sigma$ of Marr wavelet to be 1/2 of the $\sigma$ used for L4, L2/3 layers as its dimension is 1/2 times the dimension of the L4 and L2/3 layer. Therefore, the L5 activity after the lateral influences step is given by,

\begin{multline}
    L5_{i,j}^{\prime}=L5_{i,j}\left [ 1-\sum_{k,l}M_{\sigma/2}(k,l;i,j) \right ] \\ + \sum_{k,l} L5_{k,l} M_{\sigma/2}(i,j;k,l)
    \label{equ_7}
\end{multline}

\paragraph{}
Finally, for the back transition from layer 5 to layer 2/3 the activities are computed as,

\begin{equation}
    L2/3_{i,j}=\sum_{k=i-12}^{i+12} \sum_{l=j-12}^{j+12} w_3^{i,j,k,l} L5_{k,l}^{\prime}
    \label{equ_8}
\end{equation}

\begin{multline}
    L2/3_{i,j}^{\prime}=L2/3_{i,j}\left [ 1-\sum_{k,l}M_{\sigma}(k,l;i,j) \right ] \\ + \sum_{k,l} L2/3_{k,l} M_{\sigma}(i,j;k,l)
    \label{equ_9}
\end{multline}

\paragraph{}
After each 2-step process, we apply normalization to the activity values in every layer with respect to the maximum values to keep the maximum activity at 1. We also apply the necessary padding at each step.

\subsection{V1 inspired SOM model}
\paragraph{}
The self-organizing map \cite{Kohonen_som} is one of the most popular unsupervised learning algorithms for non-linear data projection for visualization. The self-organization map and its variants have many applications \cite{malek,Onishi,bezdek,xu, laha}.  Inspired by V1 lateral influences we also design a modified version of the Self-organizing map algorithm. Kohonen's Self-organizing map \cite{Kohonen_som}, though considers the lateral distance from the winner neuron, the weight update of each neuron actually happens independent of the weights  of the spatially nearby neurons. In our modified version we consider the lateral influence of neighborhood neurons in terms of weight sharing. For a SOM grid of size ($n\times n$), let $d^2(i,j;k,l)=(i-k)^2+(j-l)^2$ be the squared distance between node (i,j) and (k,l) on that grid. Weight vector $W_{i,j}\in R^p$ is the $p$-dimensional weight vector associated with the node (i,j), $p$ is the input dimension. Let (r,s) be the winner node for an input x.
Given an input x, every neuron should get a chance to update its weight. Thus for an arbitrary neuron $(i,j)$, the update to its weight $W_{i,j}$ should be influenced by both the winner as well as other neurons due to the lateral interaction between neurons. In the case of Kohonen's SOM, the update to $W_{i,j}$ depends only on the winner node. In our proposal, the update to $W_{i,j}$ not only depends on the similarity between $x$ and $W_{i,j}$ and $d^2(i,j;r,s)$ but also on spatial weights and spatial locations of the nodes $(k,l)$. In particular, updates to $W_{i,j}$ depends on $(x - W_{k,l})$ and $d^2(i,j;k,l)$. Of course, the influence of the winner remains the strongest.

\begin{figure*}[!ht]
    \centering
    \includegraphics[width=\textwidth]{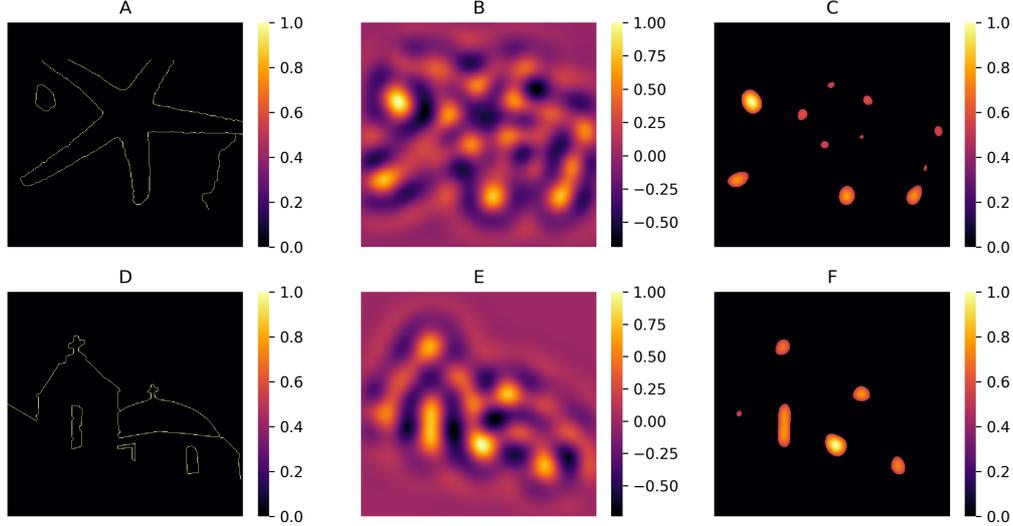}
    \caption{Examples of V1 action. \textbf{A:} and \textbf{D:} Input images (LGN); \textbf{B:} and \textbf{E:} V1 outputs; \textbf{C:} and \textbf{F:} V1 outputs with ReLu(x-0.5) applied.}
    \label{fig_3}
\end{figure*}

\begin{figure*}[!ht]
    \centering
    \includegraphics[width=0.85\textwidth]{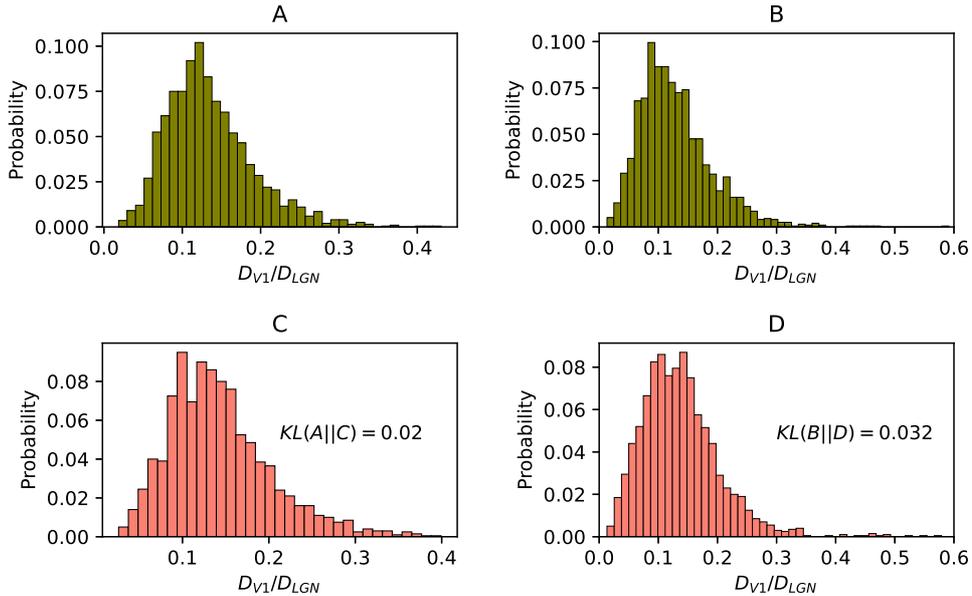}
    \caption{V1 distortion reduction. Probability distribution of the ratio $D_{V1}/D_{LGN}$ calculated for 10 distorted copies for each of the 200 inputs. \textbf{A:} ReLu(x-0.5) and \textbf{B:} ReLu(x-0.6) in presence of iso-orientation domains. \textbf{C:} ReLu(x-0.5) and \textbf{D:} ReLu(x-0.6) in absence of iso-orientation domains. The Kullback–Leibler (KL) divergence computed between distributions in \textbf{A} and \textbf{C}, $KL(A||C)$ and that between distributions in \textbf{B} and \textbf{D}, $KL(B||D)$ are also provided on the figure.}
    \label{fig_4}
\end{figure*}

\begingroup
\Large
\begin{multline}
    W_{i,j}^t = W_{i,j}^{t-1}+  
     \eta_t\, exp \left [-\frac{d^2(i,j;r,s)}{2\gamma^2_t} \right ] \\ \times \frac{\sum_{k,l} (x-W_{k,l}^{t-1})\, exp \left [-\frac{d^2(i,j;k,l)}{2\sigma^2_t} \right ]}{\sum_{k,l} exp \left [-\frac{d^2(i,j;k,l)}{2\sigma^2_t}\right ]}
     \label{equ_10}
\end{multline}
\endgroup

\paragraph{}
At epoch $t$, $\eta_t=\eta_0 e^{-K_0 t}$, $\gamma_t=\gamma_0 e^{-K_1 t}$ and $\sigma_t=\sigma_0 e^{-K_2 t}$. The first exponential factor in our update equation considers the distance from the winner, while the second exponential term considers the influence of other neurons in the vicinity, and the strength of influence gets diminished as the spatial distance of the influencing neuron increases. In this context, it is worth mentioning the study in \cite{christian}  has demonstrated the relevance of strong lateral competition in the formation of Self-organizing maps.

\section{Experiments and Results}
\subsection{V1 can extract low-level features}

\paragraph{}
We used our V1 model on the BSDS500 dataset \cite{bsds500}. We took the 200 ground truth images (by artist 3) given in this dataset as inputs for our model. We found that our V1 model can extract low-level features from the inputs having sharp high-level features. We further applied the ReLu activation function to the V1 outputs to separate out the enhanced low-level features from the V1 outputs. Figure \ref{fig_3} shows two examples of the action of our V1 model.

\subsection{V1 can reduce distortion}

\paragraph{}
Next, we added distortion to the input images by randomly interchanging each pixel value with that of one of its 8 nearest neighbors, to check for the robustness of our V1 model. To measure the amount of distortion we take the Frobenius norm of the difference between the original matrix and the distorted matrix. Therefore, the amount of distortion between an original LGN matrix and its distorted copy and between their corresponding V1 outputs are given by,

\begin{equation*}
    D_{LGN}=\sqrt{Tr[(LGN_o-LGN_d)(LGN_o-LGN_d)^T]}
\end{equation*}

\begin{equation*}
    D_{V1}=\sqrt{Tr[(V1_o-V1_d)(V1_o-V1_d)^T]}
\end{equation*}

\paragraph{}
We generated 10 different distorted copies for each of the 200 inputs and computed the value of $D_{V1}/D_{LGN}$ for every distorted input and its V1 output. Figures \ref{fig_4}(A) and (B) show the distribution of $D_{V1}/D_{LGN}$ for 2 different ReLu thresholds. As evident from the figure, V1 can reduce the amount of distortion by $8-11$ folds. Therefore, besides extracting low-level features, our V1 model can also perform a significant  reduction of distortion in the inputs.

\begin{figure*}[!ht]
    \centering
    \includegraphics[width=0.8\textwidth]{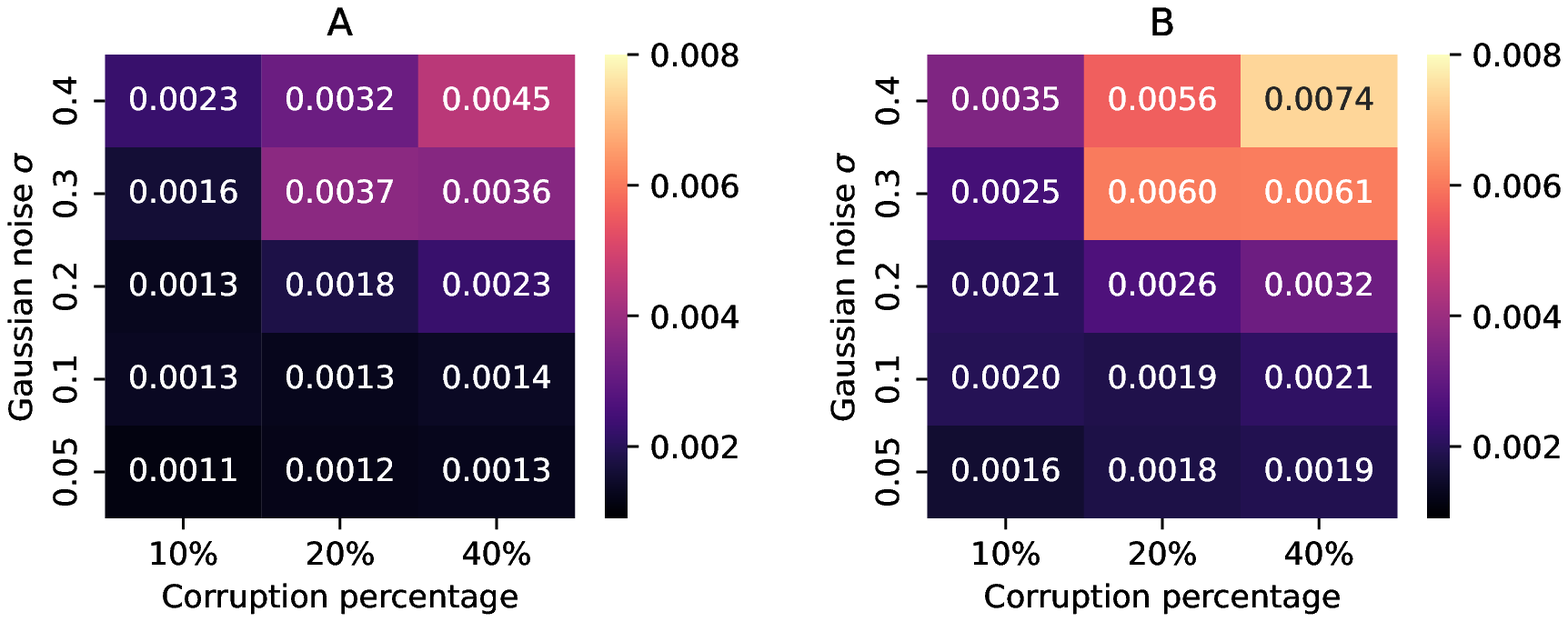}
    \caption{Quantization errors for \textbf{A:} V1-SOM and \textbf{B:} SOM at different noise levels and different $\sigma$ of Gaussian noise.}
    \label{fig_5}
\end{figure*}

\begin{figure*}[!hb]
    \centering
    \includegraphics[width=\textwidth]{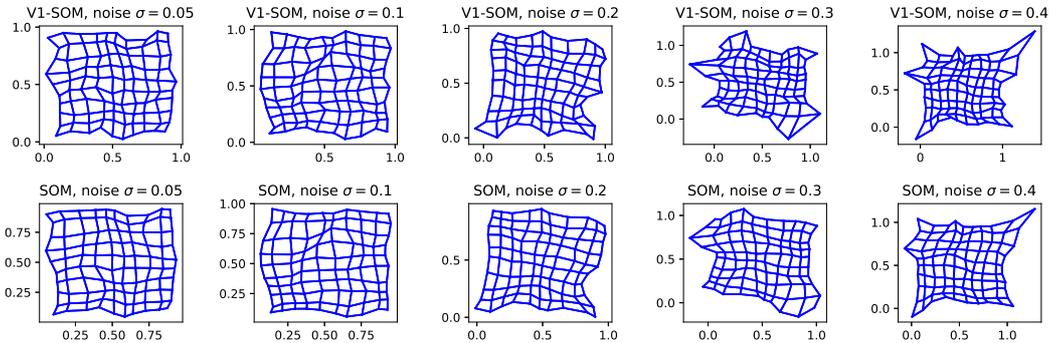}
    \caption{Weight unfolding diagrams for \textbf{Row-1:} V1-SOM and \textbf{Row-2:} SOM at $20\%$ noise level with different $\sigma$ of Gaussian noise.}
    \label{fig_6}
\end{figure*}

\begin{figure*}[!hb]
    \centering
    \includegraphics[width=0.8\textwidth]{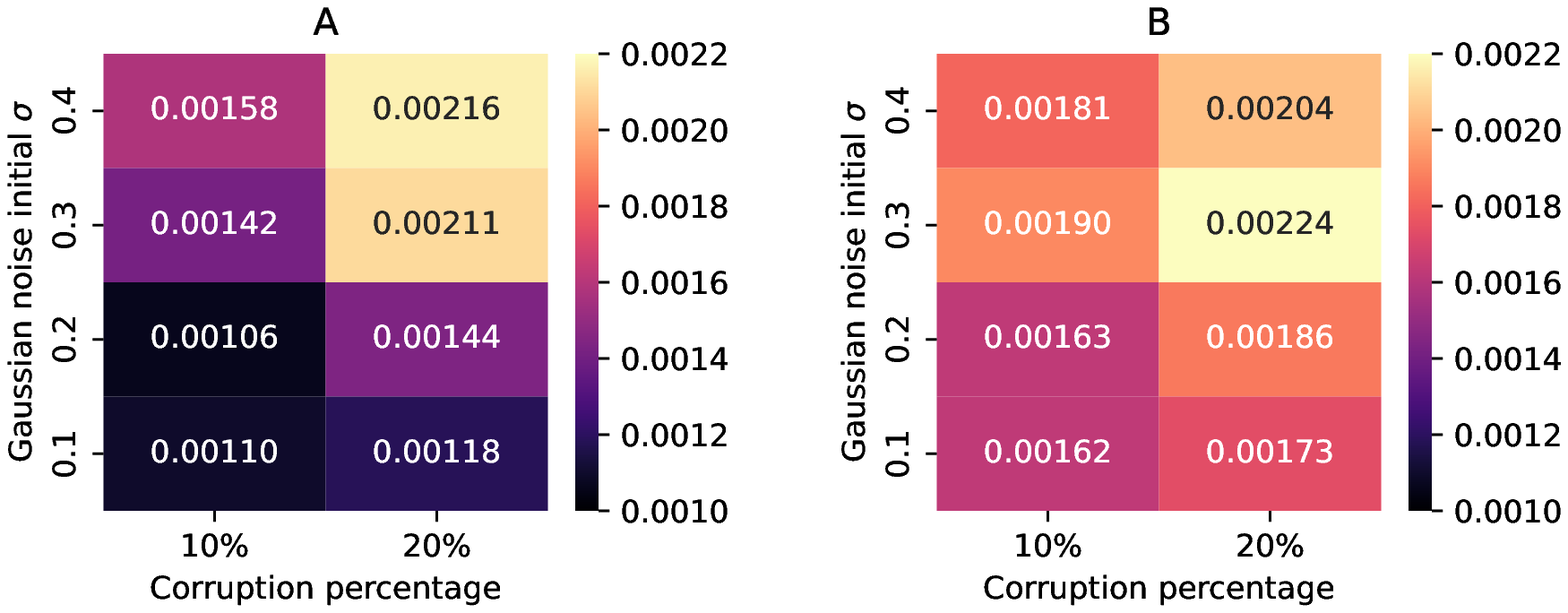}
    \caption{Quantization errors for \textbf{A:} V1-SOM and \textbf{B:} SOM at different noise levels during weight updates and different initial $\sigma$ of Gaussian noise.}
    \label{fig_7}
\end{figure*}

\begin{figure*}[!ht]
    \centering
    \includegraphics[width=0.8\textwidth]{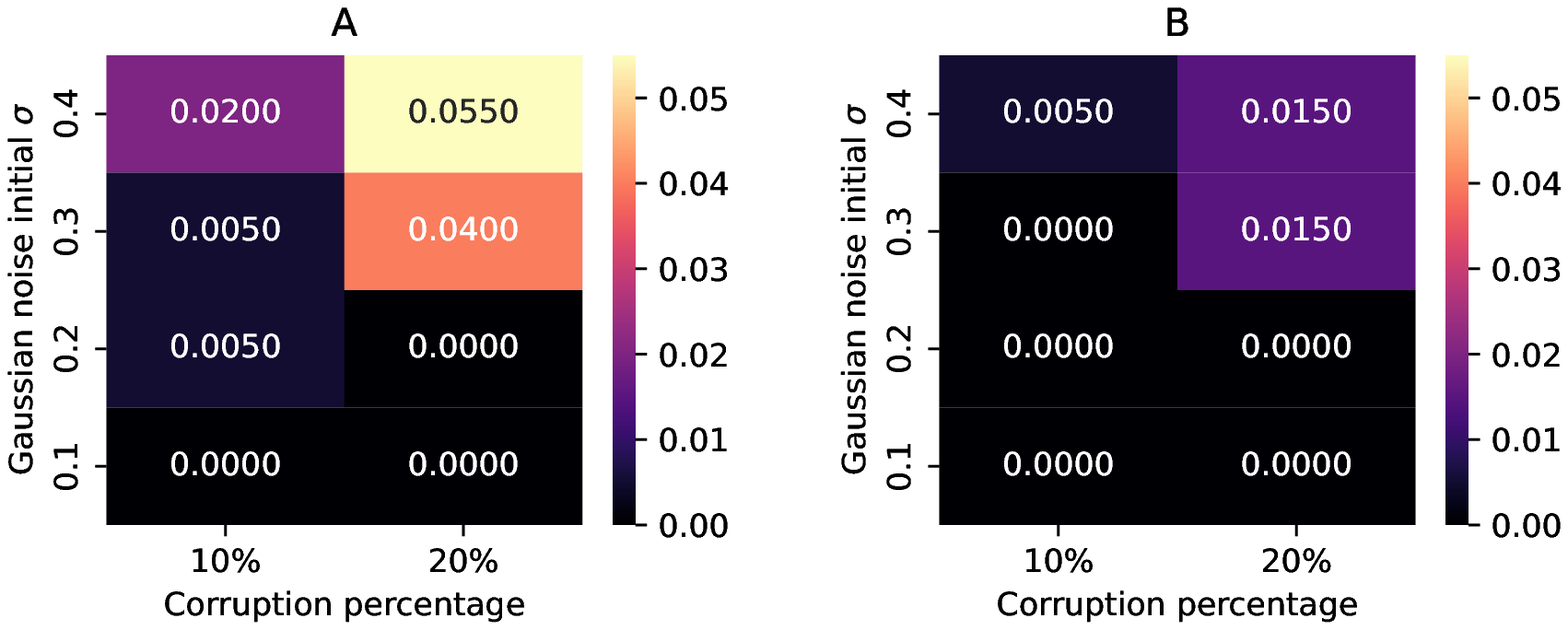}
    \caption{Topographical errors for \textbf{A:} V1-SOM and \textbf{B:} SOM at different noise levels during weight updates and different initial $\sigma$ of Gaussian noise.}
    \label{fig_8}
\end{figure*}

\subsection{Effect of iso-orientation domains}
\paragraph{}
Using our model we also wanted to check whether the presence of iso-orientation domains provides any additional benefits to V1 or not. Unlike cats and higher cognitive animals, iso-orientation domains are not found in rodents \cite{No_orientation_map_1,No_orientation_map_2,No_orientation_map_3,No_orientation_map_4,No_orientation_map_5}. In rodents different orientation-selective neurons are randomly scattered across a layer. We, therefore, made the orientation selectivity random for each neuron of each layer to replicate the structure of rodent V1. Figure-\ref{fig_4}(C) and (D) shows the distribution of $D_{V1}/D_{LGN}$ for 2 different thresholds of ReLu  for this rodent V1 model. We have also computed the Kullback–Leibler (KL) divergence between the distributions in Figures-\ref{fig_4}(A) and (C), $KL(A||C)$ and Figures-\ref{fig_4}(B) and (D), $KL(B||D)$. The KL divergence values make it clear that the presence of iso-orientation domains in V1 layers carries little additional benefits in terms of distortion reduction, which gives confirmation to the prediction made in \cite{No_orientation_map_6}. This near invariance of the results most likely occurs because  different orientation-selective neurons are randomly scattered across a layer, and consequently, $1/4^{th}$ of the neurons surrounding each neuron have the same orientation selectivity on average. Therefore, even in this case, there are multiple neurons with the same orientation selectivity within a small area, providing the same functionality of iso-orientation domains.

\subsection{V1-SOM can learn corrupted input better than SOM}
\paragraph{}
After the results of our V1 model, to test the performance of our modified self-organizing map algorithm and compare it with the traditional version of the algorithm, we ran both of them on 2D random uniform distribution ($0-1$) with additive Gaussian noise. We added 2D Gaussian noise to some percentage of points chosen at random. We varied both the percentage of corruption and the $\sigma$ of the Gaussian noise to test the efficacy of both algorithms in learning the distribution in presence of noise. We calculated the quantization and topographical errors for both of the algorithms using the formulae given in \eqref{equ_11} and \eqref{equ_12}, respectively. 

\begin{equation}
    QE=\frac{1}{N} \sum_{i=1}^N || W^i_{r,s} - x_i || 
    \label{equ_11}
\end{equation}

\begin{equation}
    TE=\frac{1}{N} \sum_{i=1}^N T(r_i,s_i;r_i^\prime,s_i^\prime)
    \label{equ_12}
\end{equation}

\paragraph{}
where the summation runs over the number of data points, $(r_i,s_i)$ and $(r_i^\prime,s_i^\prime)$ are the 1st and 2nd winner nodes for data point $x_i$, $W^i_{r,s}$ is the weight vector associated with the node $(r_i,s_i)$ and the value of the function $T=0$, when $(r_i,s_i)$ and $(r_i^\prime,s_i^\prime)$ are neighbors and $T=1$ otherwise. While calculating the quantization and topographical errors, we used the datasets including the corrupted points. Figure-\ref{fig_5} shows the heatmaps of the quantization errors with increasing levels of corruption and increasing $\sigma$ of Gaussian noise. Figure \ref{fig_5} makes it clear that in all cases quantization error is lesser when trained with V1-SOM, with the same set of parameters. Topographical error on the other hand is zero for all of the cases for both V1-SOM and SOM. Therefore, additional lateral influences considered in V1-SOM enhance the robustness of the self-organizing map algorithm in learning noisy inputs. Figure \ref{fig_6} shows the weight unfolding diagrams for V1-SOM and SOM at $20\%$ corruption level of the inputs. As evident from the figure, SOM weight unfolding diagrams are relatively  smoother, but V1-SOM weights move closer to the noisy inputs at the expense of smoothness. 

\subsection{V1-SOM can tolerate noise during weight update better than SOM}
\paragraph{}
Other than noisy inputs there can be noise during weight updates too. This is a more plausible situation for biological neural systems.  To test the efficacy of V1-SOM against noisy weight updates relative to regular SOM, we added Gaussian noise to some percentage of weights chosen at random after each epoch. However, we took the $\sigma$ of the Gaussian noise to be exponentially decaying (similar to $\sigma_t=\sigma_0 e^{-K_2 t}$), as noise with high $\sigma$ when the map tries converge will be harmful. Similar to the previous case, we varied both the percentage of weight corruption and the initial $\sigma$ of the Gaussian noise and ran both SOM and V1-SOM on 2D random uniform ($0-1$) inputs. We calculated the quantization and topographical errors for both of the algorithms. Figures \ref{fig_7} and \ref{fig_8} show the heatmaps of the quantization and topographical errors with increasing levels of weight corruption during training and increasing initial $\sigma$ of added Gaussian noise. It is evident from the figures that V1-SOM can tolerate noisy weight updates better than SOM up to $20\%$ weight corruption during every epoch and up to the initial $\sigma=0.2$ of the Gaussian noise added. Therefore, V1-SOM performs better than SOM for moderate levels and intensities of noise during weight updates.

\subsection{V1-SOM on MNIST data}

\begin{figure}[H]
    \centering
    \includegraphics[width=\linewidth]{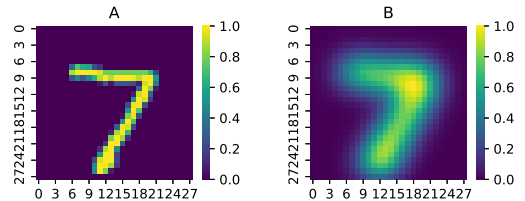}
    \caption{Example of V1 processing on MNIST data. \textbf{A:} original image and \textbf{B:} V1 processed image.}
    \label{fig_9}
\end{figure}

\begin{figure*}[!ht]
    \centering
    \includegraphics[width=\textwidth]{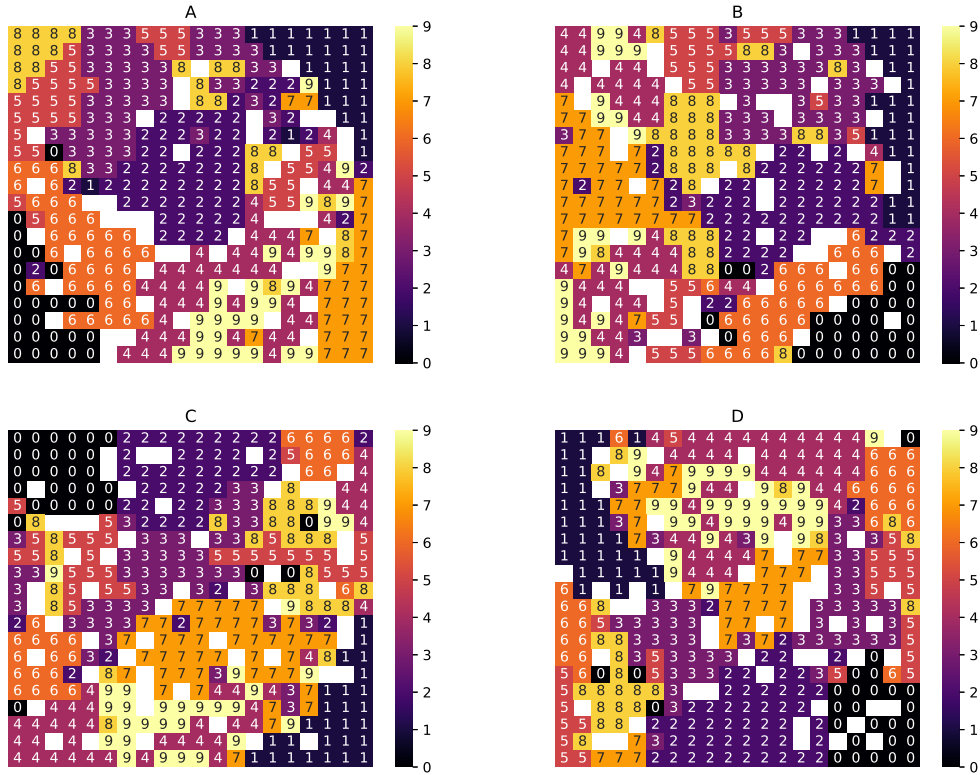}
    \caption{Final maps for \textbf{A:} V1-SOM on MNIST data; \textbf{B:} SOM on MNIST data; \textbf{C:} V1-SOM on V1 processed MNIST data and \textbf{D:} SOM on V1 processed MNIST data, starting with same initial weights.}
    \label{fig_10}
\end{figure*}

\begin{table*}[!ht]
\centering
\caption{Quantization errors}
\label{Table_1}
\resizebox{\textwidth}{!}{
\begin{tabular}{ccccc}
\hline
\rowcolor[HTML]{FFCE93} 
Trial & V1-SOM on MNIST & SOM on MNIST & V1-SOM on V1 processed MNIST & SOM on V1 processed MNIST \\ \hline
1 & 23.26           & 23.57        & 3.42                         & 3.46                      \\ \hline
2 & 23.37           & 23.58        & 3.42                         & 3.42                      \\ \hline
3 & 23.36           & 23.58        & 3.45                         & 3.44                      \\ \hline
4 & 23.44           & 23.46        & 3.43                         & 3.43                      \\ \hline
5 & 23.28           & 23.6         & 3.44                         & 3.45                      \\ \hline
\end{tabular}}
\end{table*}

\begin{table*}[!ht]
\centering
\caption{Topographical errors}
\label{Table_2}
\resizebox{\textwidth}{!}{
\begin{tabular}{ccccc}
\hline
\rowcolor[HTML]{FFCE93} 
Trial & V1-SOM on MNIST & SOM on MNIST & V1-SOM on V1 processed MNIST & SOM on V1 processed MNIST \\ \hline
1     & 0.006           & 0.008        & 0.008                        & 0.011                     \\ \hline
2     & 0.006           & 0.006        & 0.015                        & 0.007                     \\ \hline
3     & 0.006           & 0.003        & 0.013                        & 0.008                     \\ \hline
4     & 0.004           & 0.005        & 0.016                        & 0.008                     \\ \hline
5     & 0.001           & 0.007        & 0.02                         & 0.011                     \\ \hline
\end{tabular}}
\end{table*}

\paragraph{}
Finally, we apply our V1-SOM model to the MNIST dataset. We use the first 1000 out of the 10000 images, that are given in the test set of MNIST, as the training set for our model. We also generated a low-level MNIST dataset by applying V1 processing to the original MNIST dataset. However, as the dimension of the MNIST images is only (28x28) our complete V1 model cannot be applied to such small images. We, therefore, applied a simplified version of it by considering only the lateral influence step with a small $\sigma$ and applying it 4 times. An example of the low-level MNIST dataset obtained from V1 processing is shown in Figure \ref{fig_9}.

\paragraph{}
For comparison with regular SOM we applied V1-SOM and SOM on both the original and the V1-processed MNIST dataset. Tables \ref{Table_1} and \ref{Table_2} show the quantization and topographical errors for 5 different initial conditions. As evident from the tables the performance of V1-SOM is nearly the same as that of the regular SOM for this high dimensional data (dimension$=28\times28=784$). Applying V1 processing on the MNIST dataset however reduces the quantization error significantly, at the expense of a slight increase in the topographical error. This result is in accordance with the \textit{Manifold untangling hypothesis} regarding visual object recognition in the brain \cite{Manifold_hypothesis}, which states that the ventral stream ($RGC \rightarrow LGN \rightarrow V1 \rightarrow V2 \rightarrow V4 \rightarrow ITC$) performs gradual untangling of input manifolds, by pulling the inputs belonging to the same class closer and pushing the ones belonging to different classes far apart. Figure \ref{fig_10} depicts the maps produced by V1-SOM and SOM, each using both original MNIST data and the V1-processed MNIST data. The label of a node is determined by the class having the maximum number of points for which the node becomes the winner. The slight spacial delocalization of the map (Figure \ref{fig_10}), when trained with the V1 processed dataset, is also supported by experiment \cite{IT_deficiency}, where chemical inactivation of different regions in the ITC results in a deficiency in recognition of different objects. However, the deficit for a particular class of objects was not entirely tied to the inactivation of a single region, indicating a spatially diffused map for the same class of objects.

\subsection{V1-SOM on Wisconsin Breast Cancer data}

\begin{figure}[H]
    \centering
    \includegraphics[width=\linewidth]{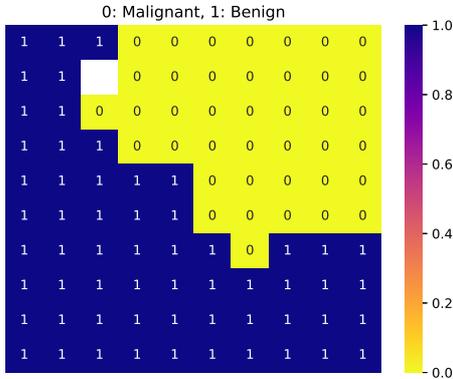}
    \caption{Map obtained by running V1-SOM on the Wisconsin Breast Cancer (Diagnostic) Dataset}
    \label{fig_11}
\end{figure}

In order to demonstrate the effectiveness of the V1-SOM for non-image data, we have run our algorithm on the Wisconsin Breast Cancer (Diagnostic) Database (WBCD) \cite{breast_cancer}, which is a 30-dimensional dataset having two classes, benign and malignant. We have run the V1-SOM algorithms 5 times on a 10x10 map. The average quantization error and the topographic error are $0.111$ and $0.08646$ respectively. Figure-\ref{fig_11} depicts a typical map for this data set, which clearly reveals the capability of V1-SOM to separate the two classes preserving the topology.


\section{Conclusion and Discussion}
\paragraph{}
We designed an approximate model of V1 which is based on the detailed anatomy of the primary visual cortex. The orientation selectivity of the V1 neurons helps in breaking up the input contour into separate fragments. In the follow-up lateral influence step, the neurons share each others' activities and as a result, the segments of the input which have a higher density of points get enhanced. This way V1 performs low-level feature extraction. In case of distorted inputs, if one neuron which detected a line segment from the original input, fails to detect the same segment from the distorted input, another neuron in the neighborhood might detect that misplaced segment from the distorted version. The chances of this happening are higher in presence of iso-orientation domains. Nevertheless, it can still happen in absence of iso-orientation domains as shown earlier. In the lateral influence step, information sharing between neighbors, therefore, causes the enhancement of nearly the same regions as with the original inputs. This way V1 can perform distortion reduction task from the inputs. 

\paragraph{}
Similar to our V1 model, in V1-SOM the weight update of each neuron gets influenced by its neighbors. This cooperative information sharing is helpful when presented with noisy data or when noise is added during weight updates, just like V1 helps in the reduction of distortion. However, 2 non-biological aspects that still remain in V1-SOM are the dependence of weight update on the distance from the winner neuron, and the use of euclidean distance-based calculations instead of dot-product-based calculations. The earlier LISSOM model \cite{Lissom} does not consider these 2 factors. However, the LISSOM model needs pre-training with regular SOM in order to work properly, which is a weak point of that model.

\paragraph{}
We are still far from a proper understanding of how the brain performs visual object recognition. In this work, we were able to shed some light on the functioning of V1 only. Similar anatomy-inspired models must be made for V2, V4, and ITC also, to get a complete understanding of how the ventral stream works. However, there have not been as many experiments done on these 3 later regions of the ventral stream, as have been done on the V1 region. For example, ITC most likely maps objects with different complex features to different clusters of neurons \cite{ITC_1} i.e. performs the self-organizing part. However, in order to design a proper self-organizing map algorithm, based on the ITC will require more detailed anatomical experiments on this region. The use of Spiking Neuron models can also be beneficial as they capture the dynamics of neuronal membrane potentials more accurately \cite{Zenke2015,Clopath2021}. We hope this work will motivate others to conduct more experiments on V2, V4, and ITC and to design computational models of these regions to get a complete understanding of how the ventral stream functions. 

\section{Data accessibility}
\paragraph{}
The main codes for the V1 and V1-SOM are available at the following \href{https://github.com/suvamroy/Codes/tree/master/V1_V1-SOM}{Github} link.

\section{Acknowledgement}
\paragraph{}
We thank Prof. Supratim Sengupta for providing us the computational facilities used in this study.

\bibliography{reference.bib}
\bibliographystyle{unsrt}

\end{multicols}

\end{document}